\documentclass[prd,twocolumn,preprintnumbers]{revtex4-1}

\usepackage{amsmath}
\usepackage{epsfig}
\usepackage{graphicx}
\usepackage{color}
\usepackage[normalem]{ulem}
 \usepackage{url}
\usepackage[breaklinks, plainpages=false, colorlinks=true, anchorcolor=cyan, linkcolor=red, citecolor=cyan, urlcolor=magenta, bookmarks=false]{hyperref}

\usepackage[caption=false]{subfig}

\begin{document}
\renewcommand{\thefigure}{\arabic{figure}}
\setcounter{figure}{0}

 \def\I{{\rm i}}
 \def\E{{\rm e}}
 \def\D{{\rm d}}

\bibliographystyle{apsrev}

\title{Non-stationary noise in gravitational wave analyses: The wavelet domain noise covariance matrix.}

\author{Neil J. Cornish}
\affiliation{eXtreme Gravity Institute, Department of Physics, Montana State University, Bozeman, Montana 59717, USA}

\begin{abstract} 
Gravitational wave detectors produce time series of the gravitational wave strain co-added with instrument noise. For evenly sampled data, such as from laser interferometers, it has been traditional to Fourier transform the data and perform analyses in the frequency domain. The motivation being that the Fourier domain noise covariance matrix will be diagonal if the noise properties are constant in time, which greatly simplifies and accelerates the analysis. However, if the noise is non-stationary this advantage is lost. It has been proposed that the time-frequency or wavelet domain is better suited for studying non-stationary noise, at least when the time variation is suitably slow, since then the wavelet domain noise covariance matrix is, to a good approximation, diagonal. Here we investigate the conditions under which the diagonal approximation is appropriate for the case of 
the Wilson-Daubechies-Meyer (WDM) wavelet packet basis, which is seeing increased use in gravitational wave data analysis. We show that so long as the noise varies slowly across a wavelet pixel, in both time {\em and} frequency, the WDM noise correlation matrix is well approximated as diagonal. The off-diagonal terms are proportional to the time and frequency derivatives of the dynamic spectral model. The same general picture should apply to other discrete wavelet transforms with wavelet filters that are suitably compact in time and frequency. Strategies for handling data with rapidly varying noise that violate these assumptions are discussed.
\end{abstract}

\maketitle

\section{Introduction}

The LIGO-Virgo detectors~\cite{aLIGO:2020wna,Virgo:2019juy} have recorded hundreds of short duration gravitational wave signals~\cite{LIGOScientific:2025slb}, most spending seconds or less in the sensitive frequency band of the detectors. Over these short time periods, it is generally a good approximation to treat the noise as stationary and Gaussian, at least once any short duration noise transients (glitches) have been removed~\cite{LIGOScientific:2019hgc}. However, as the detector sensitivities improve, especially for third generation ground based interferometers~\cite{Evans:2021gyd,ET:2025xjr}, and for the future space based LISA interferometer~\cite{LISA:2024hlh}, signals will spend days to years in the sensitive band of the detectors, and it is inevitable that the noise properties will vary on such long time scales. The traditional frequency domain analyses that are currently used to analyze LIGO-Virgo will need to be modified to account for non-stationary noise. One alternative approach is to move to the time-frequency domain using discrete wavelet transforms of the data~\cite{Klimenko:2005xv,Klimenko:2015ypf,Cornish:2020odn,Digman:2022igm,Digman:2022jmp,Pearson:2025wfd}. The potential advantage of wavelet domain analyses is that the noise correlation matrix is expected to be well approximated as diagonal, at least for noise processes that vary slowly in time. Here we investigate the structure of the wavelet domain noise covariance matrix for the particular case of the Wilson-Daubechies-Meyer (WDM) wavelet packet basis~\cite{Necula_2012}, which is seeing increased use in gravitational wave analyses. Since wavelets have compact support in time and frequency, the wavelet noise covariance matrix is guaranteed to be band-diagonal for any noise process. We show that the WDM noise covariance matrix is well approximated as diagonal so long as the noise properties vary slowly in time {\em and} frequency across each wavelet pixel. 

\section{The noise covariance matrix}

For signals in additive noise, the probability distribution for the noise process defines the likelihood function. If the noise process is zero-mean Gaussian, the likelihood is fully determined by the noise covariance matrix ${\bf C}$, the inverse of which appears in the likelihood function. By applying linear transformations to the data, it is always possible to diagonalize ${\bf C}$, or even better, transform it to the identity matrix. This transformation is known as whitening the data. The result is a special case of the Karhunen--Lo\`{e}ve (KL) theorem~\cite{KAR46,loeve} which states that a centered stochastic time series can always be decomposed into eigenfunctions with expansion coefficients that are uncorrelated random variables. For Gaussian noise, the expansion coefficients are independent Gaussian variables. For wide sense stationary (WSS) Gaussian noise, the eigenfunctions are a Fourier series. The KL theorem tells us that ${\bf C}$ can always be diagonalized by some transformation. However this is not always very useful as the transformation can be complicated. In the case of gravitational wave data analysis, the waveform models must also be transformed to the new representation, and this can be costly for general KL bases. Moreover, interpreting the noise correlation matrix in a general KL basis can be challenging. The Fourier basis is nice because fast methods have been developed to compute the transformation - Fast Fourier Transforms (FFTs) for the data, and the Stationary Phase Approximation (SPA) for the waveform models. For WSS Gaussian noise the Fourier domain noise correlation matrix is diagonal, with the diagonal entries given by the power spectral density $S(f)$, which is easy to interpret. However, the Fourier transform does not diagonalize non-stationary noise, and the Fourier domain noise covariance matrix can become dense, making it costly to invert, difficult to interpret, and expensive to use in likelihood calculations. 

A better approximation to the noise in gravitational wave detectors is that the Gaussian component (i.e. after removing any glitches) is locally stationary~\cite{LIGOScientific:2019hgc}. That is, the noise covariance varies slowly with time. The physical nature of locally stationary noise suggests that something like a short duration Fourier transform (SFT) should come close to matching the KL eigenfunctions. Mallat et. al.~\cite{10.2307/119978} investigated using short duration (windowed) Fourier bases to model locally stationary noise, and found that the noise covariance matrix was banded and diagonal dominant. The WDM wavelet basis falls into the general category considered by Mallat, but makes the complementary window choice. The WDM basis uses time-domain window functions that oscillate within an overall Gaussian envelope, leading to a smoothed top-hat window function in the frequency domain. Mallat instead considered smoothed top-hat window functions in the time domain, which result in decaying oscillations in the frequency domain. More recently, Tenorio \& Gerosa~\cite{Tenorio:2025gci} and Du, Luo \& Xu~\cite{Du:2025fes} considered SFTs as an alternative to wavelets. Tenorio \& Gerosa use a rectangular window function in the time domain, while Du, Luo \& Xu use Tukey windows in the time domain, which lessens the ringing in the frequency domain, and corresponds more closely to the Mallat et. al. approach. 

Discrete wavelet transformations quantize the time-frequency plane into pixels with time extent $\Delta T$ and frequency extent $\Delta F$ satisfying $\Delta T \Delta F = 1/2$. The most common types of transform yield either dyadic decompositions (logarithmic spacing in $\Delta F$) or binary decompositions (linear spacing in $\Delta F$). Strictly speaking, the binary decompositions are wavelet wave packets, not pure wavelets. There is a trade-off in the choice of $\Delta T$ - we want $\Delta T$ to be short compared to the timescale over which the noise properties are varying, but choosing $\Delta T$ too small leads to a large $\Delta F$, which we will see can lead to correlations between wavelet pixels at different times. 

\section{The WDM Transform}

The WDM wavelet wave packets $g_{nm}$ form a complete, orthogonal basis that can be used to faithfully represent any time series:
\begin{equation}
x[k] = \sum_{n = 0}^{N_t-1}\sum_{m=0}^{N_f-1}  w_{nm} \, g_{nm}[k] \, .
\end{equation}
A time series of duration $T$ sampled at a cadence $\Delta t = T/N$ can be represented by a rectangular grid with $N_t$ time slices of width $\Delta T$ and $N_f$ frequency slices of width $\Delta F$:
\begin{eqnarray}
\Delta T &= &N_f \, \Delta t \nonumber \\
\Delta F &= &\frac{1} { 2 \Delta t\, N_f} \, .
\end{eqnarray}
Each of the $N = N_t N_f$ cells has area $\Delta T \Delta F = 1/2$. The time and frequency resolution can be varied by changing $N_f$, such that as $N_f \rightarrow N/2$ the expansion approaches a Fourier series, and as $N_f \rightarrow 1$ the expansion approaches the original time series.  The orthogonality condition
\begin{equation}
 \sum_{k = 0}^{N-1}g_{nm}[k] g_{n'm'}[k] = \delta_{nn'} \delta_{mm'}
\end{equation}
can be used to find an expression for the expansion coefficients:
\begin{equation}\label{coeff}
w_{nm} =  \sum_{k = 1}^{N} x[k] g_{nm}[k] \, .
\end{equation}
Our goal is to compute the noise covariance matrix $S_{mn,m'n'} = {\rm E}[w_{nm} w_{n'm'}]$. For white, stationary noise we have
\begin{equation}
 {\rm E}[ x[i] x[j] ] = C_{ij} = \sigma^2 \delta_{ij}
\end{equation}
and
\begin{eqnarray}
 S_{mn,m'n'}  &=&  \sum_{k = 1}^{N}  \sum_{k' = 1}^{N} {\rm E}[ x[k] x[k']] g_{nm}[k] g_{n'm'}[k']   \nonumber \\
 &=& \sigma^2 \sum_{k = 1}^{N} g_{nm}[k] g_{n'm'}[k] \nonumber \\
 &=& \sigma^2 \delta_{nn'}\delta_{mm'} \, .
\end{eqnarray}
Here the first equality uses the white-noise covariance $C_{kk'}=\sigma^2\delta_{kk'}$, which collapses the double sum to a single sum, and the second equality follows from the orthogonality of the WDM basis.
Thus, the WDM noise correlation matrix is diagonal for stationary white noise, just as it is in the time domain and the Fourier domain. Moreover, no matter what the form of the noise covariance matrix $C_{ij}$, the matrix $S_{mn,m'n'}$ will be banded since the wavelets have finite support. Thus $S_{mn,m'n'} = 0$ for $|\Delta m| = |m-m'| > 1$ and $|\Delta n| = |n-n'| > 2q$, where $q$ sets the width of the time domain wavelet filter $T_{\rm filt} = 2 q \Delta T$.

The WDM wavelets are defined in the frequency domain by:
\begin{eqnarray}\label{wavelets}
&& {\tilde g}_{nm}(\omega) = e^{-i n \omega \Delta T} \left(\alpha_{nm}\Phi(\omega - m \Delta \Omega) \right. \nonumber \\
&&\hspace*{1.1in} \left. + \alpha^*_{nm} \Phi(\omega + m \Delta \Omega)\right)  \, ,
\end{eqnarray}
where $\Delta \Omega = 2\pi \Delta F$, $\alpha_{nm} = 1$ for $(n+m)$ even and  $\alpha_{nm} = i$ for $(n+m)$ odd and $\Phi(\omega)$ is the Meyer window function~\cite{meyer1990ondelettes}
\begin{equation} \label{phif}
\Phi(\omega) =  \left\{\begin{array}{lr}
       \frac{1}{\sqrt{\Delta \Omega}} , &  |\omega| <  A\\
         \frac{1}{\sqrt{\Delta \Omega}}\cos\left[ \frac{\pi}{2}\nu_d\left( \frac{|\omega| -A}{B}\right) \right], & A\leq |\omega| \leq A+B\\
         0, & |\omega| > A+B
        \end{array}\right. \, ,
\end{equation}
where $\nu_d(x)$ is the unit-normalized incomplete Beta function,
\begin{equation}
\nu_d(x) = I_x(d+1,d+1) = \frac{1}{B(d+1,d+1)}\int_0^x t^d(1-t)^d\,dt
\end{equation}
for $0\leq x\leq 1$, with $\nu_d(x)=0$ for $x<0$ and $\nu_d(x)=1$ for $x>1$. The quantities $A$, $B$ satisfy the constraint $2A+B = \Delta \Omega$. The choice of $A$ and $d$ set the shape of the frequency domain and time domain window functions. One of the nice features of the WDM wavelet basis is that they can be computed at cost $N\log N$ using fast Fourier transforms. In the expressions below we focus on the interior frequency layers, $1\leq m \leq N_f-1$. The DC and Nyquist layers have half-width frequency pixels and only one valid time parity, and are omitted here to avoid obscuring the main sign conventions.

Starting with time domain data $x[i]$, the transform is given by
\begin{equation}\label{fullt}
w_{nm} =  {\cal A}_t \Re\, \alpha_{nm} \sum_{k = -K/2}^{K/2-1}  e^{i\frac{\pi k m}{N_f}} x[n N_f + k]\phi[k].
\end{equation}
with $K= 2 q N_f$, $\alpha_{nm}=1$ for $(n+m)$ even, $\alpha_{nm}=i$ for $(n+m)$ odd, and ${\cal A}_t$ is the normalization used to make the transform orthonormal. Equivalently, for $(n+m)$ even this is a cosine transform, while for $(n+m)$ odd it is minus the sine transform. The time-domain window $\phi(t)$ is given by the inverse Fourier transform of $\Phi(\omega)$.

Starting with frequency domain data $\tilde{x}[k]$, we first form
\begin{equation}
Z_{nm}= \sum_{l = -N_t/2}^{N_t/2-1}
 e^{2\pi i (l+N_t/2)n/N_t}
 \tilde{x}\left[l+\frac{mN_t}{2}\right]\Phi[l] \, .
\end{equation}
The WDM coefficients are then
\begin{equation}\label{fullf}
w_{nm} =  \left\{\begin{array}{ll}
{\cal A}_f \Re Z_{nm}, & (n+m)\;{\rm even} \\
(-1)^m{\cal A}_f \Im Z_{nm}, & (n+m)\;{\rm odd}
\end{array}\right. \, .
\end{equation}
The normalization ${\cal A}_f$ is fixed by Parseval's theorem. The total power is given by
\begin{equation}
 P =  \sum_{m=0}^{N_f-1} \sum_{n=0}^{N_t-1}  w^2_{nm}
\end{equation}
Using the frequency domain expression (\ref{fullf}), the sum over $n$ produces Kronecker deltas from the $e^{\frac{2\pi i (l-j) n}{N_t}}$ terms, leading to the expression
\begin{equation}
P = \Delta \Omega  \sum_{m=0}^{N_f-1}  \sum_{l = -N_t/2}^{N_t/2-1}   \vert \tilde{x}[l+\frac{mN_t}{2}] \vert^2 \Phi^2[l]
\end{equation}
We recognize this expression as the sum of the Fourier power in overlapping frequency bands, multiplied by the window function in each band. The square of the window function (\ref{phif}) is flat across part of the band and equal to $1/ \Delta \Omega$, then it drops to zero via a squared cosine taper with argument $\theta(\omega) =  \frac{\pi}{2}\nu_d\left( \frac{|\omega| -A}{B}\right)$. In the region where one window function is dropping to zero, the other is rising from zero via the same cosine taper. Using $\nu_d(x)+\nu_d(1-x) = 1$, we see that the adjacent windows have the complementary argument $\pi/2 - \theta(\omega)$ in the overlap region. Since $\cos^2(\pi/2 - \theta(\omega)) = \sin^2\theta(\omega)$, the sum of the rising and falling squared window functions in each overlap region is equal to $1/ \Delta \Omega$.
Thus we have
\begin{equation}\label{parseval}
 P =  \sum_{m=0}^{N_f-1} \sum_{n=0}^{N_t-1}  w^2_{nm} =  \sum_{k=0}^{N/2}    \vert \tilde{x}[k] \vert^2 \, ,
\end{equation}
which proves Parseval's theorem for the WDM wavelet basis once the discrete Fourier-transform normalization is included in ${\cal A}_f$.

We can gain insight into the structure of $S_{mn,m'n'}$ for the general case of non-stationary, colored noise by studying two limiting cases that are analytically tractable: uncorrelated non-stationary noise, where $C_{ij} = \delta_{ij} \sigma^2_i$; and wide sense stationary (WSS) colored noise, where $C_{ij} = C(|i-j|)$. Extrapolating from these limiting cases, it will be argued that the WDM noise correlation matrix is approximately diagonal for noise processes that vary slowly in time and frequency, where here the relevant time scale is the temporal extent of the pixels $\Delta T$ and the relevant frequency scale is the bandwidth of the pixels $\Delta F$. So long as the fractional change in the dynamic spectrum $S(f,t)$ across a wavelet pixel is less than $\sim 10\%$, the off diagonal elements of the WDM noise correlation matrix are of order 1\% or less. Moreover, the WDM noise correlation is banded around the diagonal no matter how fast $S(f,t)$ varies, with the correlations vanishing for times and frequencies separated by the filter width in time or frequency.

\subsection{Wide Sense Stationary Noise}

The WDM transform can be efficiently computed by first Fourier transforming the data: $x[k] \rightarrow \tilde x[j]$. Using the frequency-domain convention above, the expansion coefficients for interior layers can be written as
\begin{eqnarray}
&& w_{nm} =  \frac{{\cal A}_f}{2}  \sum_{l = -N_t/2}^{N_t/2-1}  \left( e^{2\pi i (l+N_t/2)n/N_t} \tilde{x}\left[l+\frac{mN_t}{2}\right] \right. \nonumber \\
&&\quad \quad  \left. +  e^{-2\pi i (l+N_t/2)n/N_t} \tilde{x}^*\left[l+\frac{mN_t}{2}\right]\right)\Phi[l], \; (n+m) \; {\rm even}
\end{eqnarray}
and
\begin{eqnarray}
&& w_{nm} =  \frac{(-1)^m{\cal A}_f}{2i}  \sum_{l = -N_t/2}^{N_t/2-1}  \left( e^{2\pi i (l+N_t/2)n/N_t} \tilde{x}\left[l+\frac{mN_t}{2}\right] \right. \nonumber \\
&&\quad \quad  \left. -  e^{-2\pi i (l+N_t/2)n/N_t} \tilde{x}^*\left[l+\frac{mN_t}{2}\right]\right)\Phi[l], \; (n+m) \; {\rm odd} \, .
\end{eqnarray}

For WSS noise we have
\begin{equation}\label{fcorr}
E\left[ \tilde{x}[j] \tilde{x}^*[k] \right]= \frac{1}{2} \delta_{jk} S[j] \, .
\end{equation}
Note that terms such as 
\begin{equation}
E\left[ \tilde{x}[j] \tilde{x}[k] \right]= E\left[ \tilde{x}[j] \tilde{x}^*[-k] \right] = \frac{1}{2} \delta_{j (-k)} S[j] \, ,
\end{equation}
do not contribute to the correlations since $\Phi[l]$ vanishes for $l \not\in [-N_t/2,N_t/2]$. Further note that when computing ${\rm E}[w_{nm} w_{n'm'}]$ we get terms like
${\rm E}\left[ \tilde{x}\left[l+\frac{mN_t}{2}\right] \tilde{x}^*\left[j+\frac{m'N_t}{2}\right]\right]$, which yield the condition $j = l + \Delta m N_t/2$, where $\Delta m = m-m'$ is the separation in terms of frequency layers. For $|\Delta m| > 1$ the compact support of the frequency window $\Phi[l]$ tells us there are no correlations between wavelets that are two or more frequency layers apart. 

The calculation breaks into two cases, one where $n+m$ and $n'+m'$ are both even or both odd, and another where $n+m$ and $n'+m'$ are a mix of even and odd. With the Fourier transform convention above, the signs are most transparent if we work directly with the real and imaginary parts of $Z_{nm}$. As a reminder, only the $\Delta m =0$ and $\Delta m = \pm 1$ terms are potentially non-zero. Defining $\Delta n=n'-n$ and
\begin{equation}
\Theta_l = \frac{2\pi l \Delta n}{N_t}+\pi(\Delta n+\Delta m n') \, ,
\end{equation}
and
\begin{equation}
{\cal S}_{l}^{m,\Delta m} =
S\left[l+\frac{mN_t}{2}\right]\Phi[l]
\Phi\left[l+\frac{\Delta m N_t}{2}\right] \, ,
\end{equation}
the general WDM covariance expressions for the interior layers are
\begin{eqnarray}
S_{mn,m'n'} &=& \frac{1}{2}
\sum_{l = -N_t/2}^{N_t/2-1}
\cos\Theta_l \, {\cal S}_{l}^{m,\Delta m} \, ,
\end{eqnarray}
when $n+m$ and $n'+m'$ are both even, and
\begin{eqnarray}
S_{mn,m'n'} &=& \frac{(-1)^{m+m'}}{2}
\sum_{l = -N_t/2}^{N_t/2-1}
\cos\Theta_l \, {\cal S}_{l}^{m,\Delta m} \, ,
\end{eqnarray}
when $n+m$ and $n'+m'$ are both odd. For the mixed-parity cases we have
\begin{eqnarray}
S_{mn,m'n'} &=& \frac{(-1)^{m'}}{2}
\sum_{l = -N_t/2}^{N_t/2-1}
\sin\Theta_l \, {\cal S}_{l}^{m,\Delta m} \, ,
\end{eqnarray}
when $n+m$ is even and $n'+m'$ is odd, while
\begin{eqnarray}
S_{mn,m'n'} &=& -\frac{(-1)^m}{2}
\sum_{l = -N_t/2}^{N_t/2-1}
\sin\Theta_l \, {\cal S}_{l}^{m,\Delta m} \, ,
\end{eqnarray}
when $n+m$ is odd and $n'+m'$ is even. These expressions reduce to the simpler same-layer formulas below when $\Delta m=0$.

One important case to consider is when $n=n'$ and $m=m'$. Then we have
\begin{equation}
S_{mn,mn} = \frac{1}{2}  \sum_{l = -N_t/2}^{N_t/2-1}  S\left[l+\frac{mN_t}{2}\right] \Phi^2[l] \nonumber \\
\end{equation}
This tells us that the power in each wavelet pixel is equal to the sum of the Fourier power across the pixel, convolved with the square of the frequency domain window function. The latter is basically a top-hat window with width $\Delta F$. Another interesting case is when $\Delta m=0$, where the temporal correlations between pixels are then given for $|\Delta n|$ even by
\begin{equation}
S_{mn,mn'} = \frac{1}{2}  \sum_{l = -N_t/2}^{N_t/2-1} \cos\left(\frac{2 \pi l \Delta n}{N_t}\right) S\left[l+\frac{mN_t}{2}\right] \Phi^2[l]\, ,
\end{equation}
and for $|\Delta n|$ odd by
\begin{equation}
S_{mn,mn'} = \frac{(-1)^{n+1}}{2}  \sum_{l = -N_t/2}^{N_t/2-1} \sin\left(\frac{2 \pi l \Delta n}{N_t}\right) S\left[l+\frac{mN_t}{2}\right] \Phi^2[l]\, .
\end{equation}
If we define the windowed power spectrum $S_m[j] =  S[j] \Phi^2[j-m N_t/2]$, these can be re-written for $|\Delta n|$ even as
\begin{equation}\label{same}
S_{mn,mn'} = \frac{1}{2}  \sum_{j=0}^{N/2} \cos\left(\frac{2 \pi j \Delta n N_f}{N}\right) S_m[j] 
\end{equation}
 and for $|\Delta n|$ odd as
\begin{equation}\label{different}
S_{mn,mn'} = \frac{(-1)^{n+1}}{2}  \sum_{j=0}^{N/2} \sin\left(\frac{2 \pi j \Delta n N_f}{N}\right) S_m[j] 
\end{equation}
We recognize the quantity
\begin{equation}
C_m[k] = \sum_{j=0}^{N/2} \cos\left(\frac{2 \pi j k}{N}\right) S_m[j] 
\end{equation}
as the cosine transform of the windowed power spectrum (and similarly the sine transform when $\Delta n$ is odd). Thus we have an analog of the Wiener-Kinchin theorem that relates the temporal correlation between wavelet pixels to the Fourier transform of the windowed power spectrum. However there are some key differences - one being that the correlations are only evaluated at times $T_k= \Delta n \Delta T$, the other being that it is the transform of the windowed power spectrum, and not the full power spectrum. It is because the power spectrum is windowed that the correlations between pixels at different times will turn out to be small.

Returning to the original expressions for the temporal correlations in a layer, (\ref{same}) and (\ref{different}), and evaluating them for the case where the power spectrum is constant across the region where the sum is being evaluated (corresponding to frequencies $f\in [ (m-1) \Delta F, (m+1) \Delta F]$) we find that $S_{mn,mn'} = \delta_{n n'} S_{mn,mn}$. In other words, if the power spectrum is locally flat, there are no correlations between pixels in a given frequency layer. The same also turns out to be true for adjacent layers with $\Delta m = \pm 1$. Thus, the WDM correlation matrix is perfectly diagonal for WSS noise with locally flat Fourier power spectra. 

We can go beyond the locally flat noise case by considering the Taylor expansion of the power spectrum about $f=f_0$:
\begin{equation}
S(f) = S(f_0) + S'(f_0) (f-f_0) + \frac{1}{2} S''(f_0) (f-f_0)^2 + \dots
\end{equation}
More specifically, we are interested in the spectrum expanded about the center of a wavelet frequency layer, $f_m = m \Delta F$. Note that the expansion only has to cover the region
$f\in [ (m-1) \Delta F, (m+1) \Delta F]$, so we can write
\begin{equation}
S(f) = S(f_m) \sum_{k=0}^\infty \frac{s_k}{k!}\delta f^k 
\end{equation}
where
\begin{equation}
\delta f =  \frac{(f-f_m)}{\Delta F}
\end{equation}
and $s_k= \Delta F^k \frac{d^k S}{df^k}/S$. The first few are given by
\begin{eqnarray}
s_0 &= &  1 \nonumber \\
s_1 &= &  \Delta F  (\ln S(f_m))' \nonumber \\
s_2 &=&   \Delta F^2  (\ln S(f_m))'' +s_1^2
\end{eqnarray} 
The quantity $\delta f$ ranges from -1 to 1. The quantities $s_k$ describe the dimensionless logarithmic slope of the power spectrum across a wavelet pixel. Note that $|s_1| = 1$ corresponds to a 100\% change in the power spectrum across a single wavelet pixel. In practice, for reasonable choices of the wavelet parameters, we will have $|s_k| \ll 1$ for $k \geq 1$.

To get a sense of how large the off-diagonal terms in the noise correlation matrix can be, a numerical evaluation was performed using various choices for the Meyer window function parameters $d,A,B$. The results changed by a just a few tens of percent across a wide range of values. Recent studies~\cite{Pearson:2025wfd} have adopted  the values $d=6, A = 0$ and $B = \Delta \Omega$, and for that choice we find
\begin{eqnarray}\label{freq}
&&S_{mn,mn} = X\left(1 + 0.046 \, s_2 +  0.00071 \, s_4 \right) +\dots  \nonumber \\
&& S_{mn,m(n\pm 1)} = \pm(-1)^{n+1}X  \left(0.21 \, s_1 +  0.0058 \, s_3 \right) +\dots \nonumber \\
&& S_{mn,m(n\pm 2)} = -X \left(0.029 \, s_2 +  0.00056 \, s_4 \right) +\dots \nonumber \\
&& S_{mn,(m\pm 1)(n\pm 1)} =(-1)^m X \left( \pm 0.0059 \, s_1 -   0.0030\, s_2 \right) +\dots \nonumber \\
&& S_{mn,(m\pm 1)n} = 0.
\end{eqnarray}
where $X= N_t S(f_m) /(4 \Delta  \Omega)$. In the $m(n\pm1)$ line the upper sign is for $n+1$ and the lower sign is for $n-1$. In the adjacent-frequency line the sign inside the parentheses follows $m\pm1$, while the leading $(-1)^m$ is independent of whether the neighboring time pixel is $n+1$ or $n-1$. Pixels with larger separations in time ($|\Delta n| > 2$) have smaller correlations that decrease rapidly with increasing $|\Delta n|$. By far the largest correlations occur between adjacent time pixels ($|\Delta n|=1$) in the same frequency layer ($\Delta m =0$). For $s_1=1$, these would amount to $\sim 20$\% correlations relative to the diagonal entries. But $s_1=1$ corresponds to a 100\% change in the power spectrum across a frequency layer, going from $S(f_m)/2$ to $3S(f_m)/2$ (or vice-versa, depending on the sign of $s_1$). In most cases $s_1 \ll 1$ and the correlations will be small. 

 For a power law spectrum $S(f) = A f^\alpha$ we have $s_1 = \alpha/m$ and $s_2= \alpha (\alpha -1)/m^2$ for the $m^{\rm th}$ frequency band. Frequency layers with $m < |\alpha|$ will have significant off-diagonal terms in the wavelet noise correlation matrix. These correlations can be removed by pre-whitening the data in the frequency domain using some reference spectrum, then applying the WDM transform to the pre-whitened data. 
  
 \subsection{Uncorrelated Non-Stationary Noise}
 
 The WDM transform can be efficiently computed in the time domain using the Fourier cosine and sine transformations of the windowed data. With the same convention as Eq.~(\ref{fullt}), for $(n+m)$ even 
\begin{equation}
w_{mn} = {\cal A}_t \sum_{k = -K/2}^{K/2} \cos\left( \frac{\pi k m}{N_f} \right) x[k+n N_f] \phi[k] \, ,
\end{equation}
and for $(n+m)$ odd 
\begin{equation}
w_{mn} = -{\cal A}_t \sum_{k = -K/2}^{K/2} \sin\left( \frac{\pi k m}{N_f} \right) x[k+n N_f] \phi[k] \, .
\end{equation}
 Here $K = 2 q N_f$ is the width of the window function. Later we will need the norm of the window function
 \begin{equation}
\Lambda = \sum_{k = -K/2}^{K/2}  \phi^2[k] \, .
\end{equation}

Uncorrelated non-stationary noise has the time domain covariance matrix $E[x[i]x[j]] = \delta_{ij} \sigma^2[i] $. It is useful to define the oscillatory factor,
\begin{equation}
B_{nm}(k)= \left\{\begin{array}{ll}
\cos(\pi k m/N_f), & (n+m)\;{\rm even} \\
-\sin(\pi k m/N_f), & (n+m)\;{\rm odd}\, .
\end{array}\right.
\end{equation}
The covariance can then be written as
\begin{eqnarray}
S_{nmn'm'} &=& {\cal A}_t^2 \sum_{k = -K/2}^{K/2} B_{nm}(k) B_{n'm'}(k-\Delta n N_f) \nonumber \\
 && \times \sigma^2[k+n N_f]  \phi[k] \phi[k-\Delta n N_f] \, ,
\end{eqnarray}
where $\Delta n=n'-n$. This expression keeps the signs associated with the cosine/minus-sine convention explicit. The rapidly oscillating terms involving $(m+m')$ average to zero, leaving the slowly varying terms involving $\Delta m=m'-m$.

Proceeding as we did for the frequency domain case and Taylor expanding $\sigma^2(t)$:
\begin{equation}
\sigma^2(t) = \sigma^2(t_n) \sum_{k=0}^\infty \frac{\mu_k}{k!}\delta t^k 
\end{equation}
where $t_n = n \Delta T$, and
\begin{equation}
\delta t =  \frac{(t-t_n)}{\Delta T}
\end{equation}
and $\mu_k= \Delta T^k \frac{d^k \sigma^2}{dt^k}/\sigma^2$. Let $\epsilon=+1$ for the upper frequency neighbor $m+1$ and $\epsilon=-1$ for the lower frequency neighbor $m-1$, and let $\tau=+1$ for $n+r$ and $\tau=-1$ for $n-r$. For $d=6, A = 0$ and $B = \Delta \Omega$ the largest terms in the WDM noise correlation matrix are:
\begin{eqnarray}\label{time}
&&S_{mn,mn} = Y \left(1 + 0.25 \, \mu_2 +  0.076 \, \mu_4  +0.027\,\mu_6+\dots \right) \nonumber \\
&& S_{mn,m(n\pm 2)} =  Y \left( -0.20 \, \mu_2 -\tau 0.20 \, \mu_3  -0.18\,\mu_4+\dots \right) \nonumber \\
&& S_{mn,m(n\pm 4)} =  Y \left( 0.10 \, \mu_2 +\tau 0.20 \, \mu_3  +0.27\,\mu_4+\dots \right) \nonumber \\
&& S_{mn,m(n\pm 6)} =  Y \left( -0.027 \, \mu_2 -\tau 0.082 \, \mu_3  -0.15\,\mu_4+\dots \right) \nonumber \\
&& S_{mn,(m\pm 1)n} = -\epsilon(-1)^{n+m}Y   \left( 0.25 \, \mu_1 + 0.046 \, \mu_3  +\dots \right) \nonumber \\
&& S_{mn,(m\pm 1)(n\pm 1)} =  (-1)^m Y \left( \tau 0.23\, \mu_1 +   0.11 \, \mu_2 +\dots \right)\nonumber \\
&& S_{mn,(m\pm 1)(n\pm 2)} =   \epsilon(-1)^{n+m}Y \left( 0.17\, \mu_1 + \tau 0.17 \, \mu_2  +\dots \right) \nonumber \\
&& S_{mn,(m\pm 1)(n\pm 3)} =  (-1)^m Y \left( -\tau 0.10\, \mu_1 -   0.15 \, \mu_2  +\dots \right)\nonumber \\
&& S_{mn,(m\pm 1)(n\pm 4)} =  -\epsilon(-1)^{n+m}Y \left( 0.044\, \mu_1 + \tau 0.089\, \mu_2  +\dots  \right), \nonumber \\
&& S_{mn,(m\pm 1)(n\pm 5)} =   (-1)^m  Y \left( \tau 0.012\, \mu_1 +   0.030\, \mu_2  +\dots  \right) \nonumber \\
&& S_{mn,(m\pm 1)(n\pm 6)} =     -\epsilon(-1)^{n+m}Y \left( 0.0005\, \mu_1 + \tau 0.0015\, \mu_2  +\dots  \right) 
\end{eqnarray}
where $Y = {\cal A}_t^2\Lambda\sigma^2(t_n)/2$, which reduces to $Y=\Lambda\Delta t^2\sigma^2(t_n)$ when ${\cal A}_t=\sqrt{2}\Delta t$. Terms with $\Delta m =0$ and $\Delta n$ odd vanish.
Overall, the correlations are a little larger than for the stationary colored noise case, which can be attributed to the wavelet filter function extending over $2q$ wavelet pixels in the time domain, versus 3 pixels in the frequency domain. In the above example we set $q=8$. The larger time extent of the window function is also the reason why the correlations fall off more slowly in the time direction ($\Delta n$) in this case. Nevertheless, so long as the percentage change in the noise across a wavelet pixel is less than a few percent, the off-diagonal terms are small. Note that even if the noise changes by a factor of 2 across the full extent of the wavelet filter, the change across a pixel will be $\sim 10$\%, ensuring that the off diagonal terms are small. In summary, the wavelet parameters should be chosen so that the wavelet filter length is shorter than the timescale of the non-stationarity.

\begin{figure}[htp]
\includegraphics[width=0.48\textwidth]{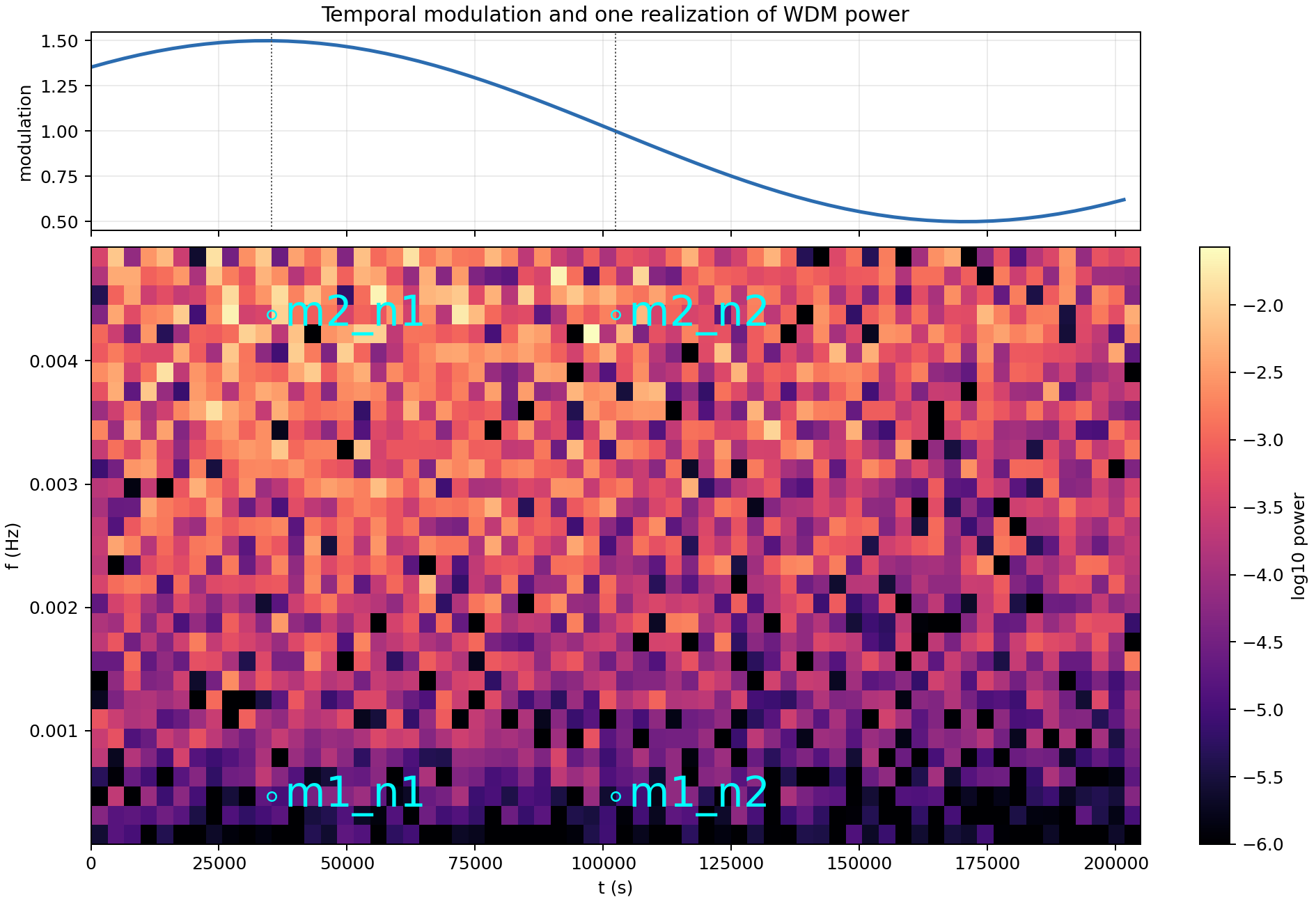} 
\caption{\label{fig:ns} The WDM power map of non-stationary noise described by a power law frequency distribution with a sinusoidally modulated power level. The upper panel shows the time modulation, while the lower panel is the power distribution across the WDM pixels for one realization of the noise process. The locations of the four pixels used to investigate the correlation are labeled with $m_1$, $n_1$, etc.}
\end{figure}

To investigate the structure of the WDM noise correlations for colored non-stationary noise, a dynamic power spectrum of the form
\begin{equation}
S(f,t) = S_0 f^\alpha \left( 1 + A \cos( 2 \pi t/\tau + \phi_0) \right)
\end{equation}
was used to generate non-stationary noise. The simulate a LISA-like example, the constants were set such that $\alpha = 2$, $A=0.5$ and $\tau = 3.16$ days. The simulated observation time covered $2.37$ days with a sample rate of 100 seconds. The low sample rate is sufficient since we focus on the low frequency region where the gradients of the power spectrum are large. The wavelet pixels had duration $\Delta T=3200$ s and bandwidth $\Delta F = 1.5625 \times 10^{-4}$ Hz. Figure~\ref{fig:ns} shows the modulation pattern and the power distribution in the WDM pixels for one noise realization.

\begin{figure}[htp]
\includegraphics[width=0.48\textwidth]{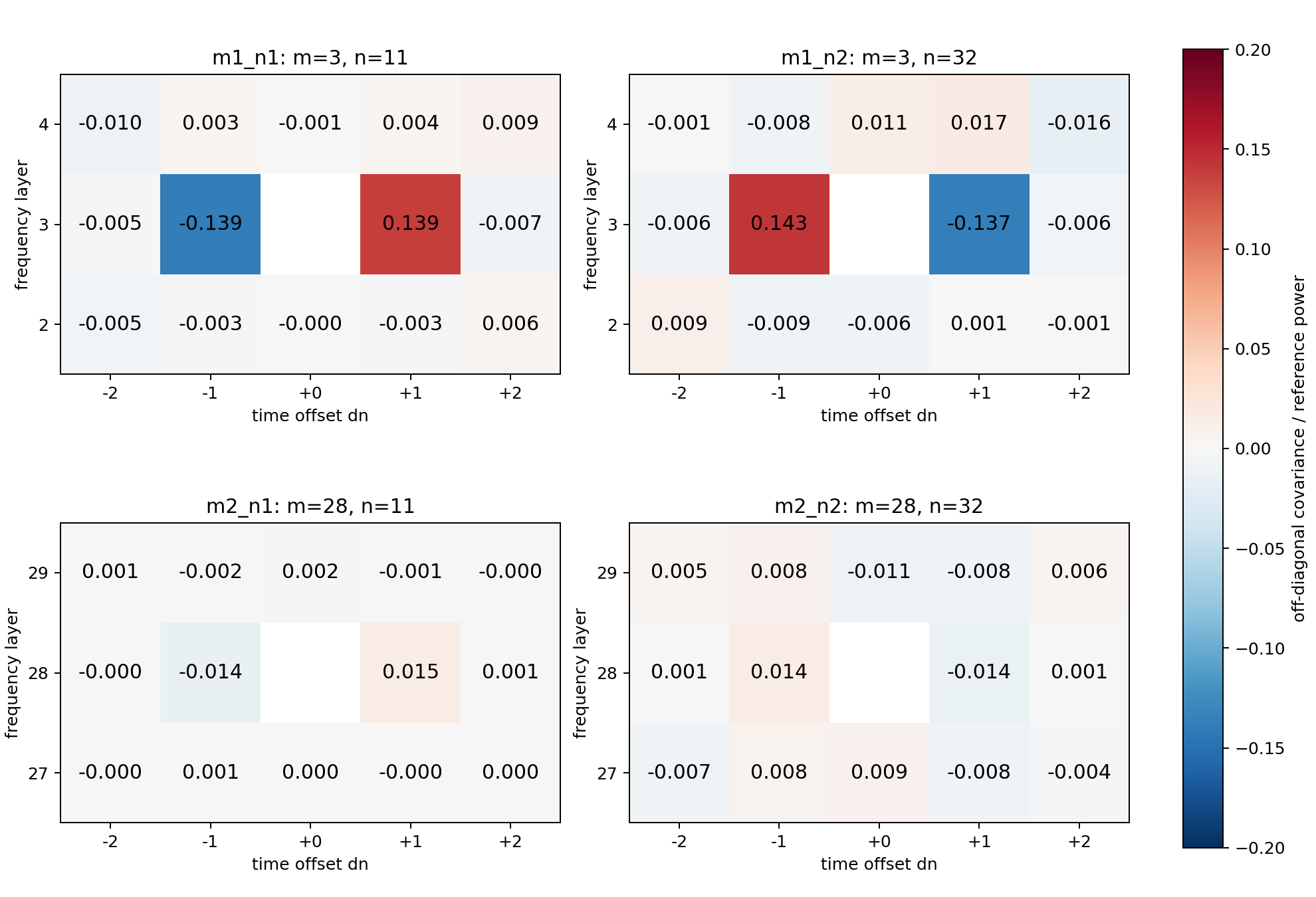} 
\caption{\label{fig:corr} The noise correlations in the neighborhoods of the four reference pixels. As expected, the low frequency pixel layer $m_1$ has the largest correlations. The times $n_2$ where the temporal modulation is changing most rapidly also have slightly larger correlations. The Monte-Carlo error on these estimates is $0.001$.}
\end{figure}

One million noise realizations we used to produce Monte Carlo estimates of the WDM correlation matrix at four representative points. The points were chosen at a low and high frequency layer, with the expectation that the correlations would be largest in the low frequency layers where the gradients in frequency are largest. Similarly, the locations of pixels in time were chosen to lie at locations where the time modulation was slowest and fastest, with the expectation that the pixel in the region with fast changing modulation would have the largest correlations. Figure~\ref{fig:corr} shows the correlation structure between the reference pixels and their nearest neighbors. The scales of the correlations are consistent with the predictions from the power-law stationary spectrum case and the uncorrelated non-stationary noise cases, which supports the hypothesis that the general $S(f,t)$ case will be very similar to the two simplified examples that allowed for an analytic treatment.

\section{Less ideal noise}

There are several scenarios that can violate the conditions needed for an approximately diagonal WDM noise covariance matrix. Two obvious examples are the sharp, high amplitude spectral lines and loud noise transients or glitches seen in LIGO/Virgo data. A typical spectral line leads to orders of magnitude changes in the power spectrum across a wavelet pixel, as do many types of glitches. The solution is to first subtract the offending feature. Spectral line subtraction has been successfully applied to LIGO/Virgo data~\cite{Sintes:1998bu,Finn:2000tt,Kimpson:2024hmk,PhysRevD.91.084034,Siegel:2024jqd}. The next generation of the {\tt BayesWave} algorithm~\cite{Cornish:2014kda,PhysRevD.91.084034,Cornish:2020dwh,Gupta:2023jrn} will feature complete line removal~\cite{cornish2025}, in addition to glitch removal. 

In the case of WSS colored noise, we saw that noise processes with very steep power spectra can lead to correlations in the lowest frequency wavelet layers. There are several ways to account for this. One is to adopt the noise subtraction techniques from pulsar timing, whereby colored noise is subtracted from the data, rather than whitened by the inverse amplitude spectral density~\cite{Lentati:2012xb,Laal:2024hdc,Gundersen:2024qmq}. Alternatively, we can first Fourier transform the data and partially whiten using the time average of $S(f,t)$, then apply the WDM transform to the partially whitened data. The pre-whitening reduces the gradients across the frequency layers, and yields a noise covariance matrix that is very close to diagonal. There is little additional cost to the pre-whitening since the fastest version of the WDM already starts with the Fourier transform of the data or signal. Similar strategies could be used to pre-process the data in the time domain if the noise is highly non-stationary.

In some applications, such as for high signal-to-noise signals in third generation detectors, it might be necessary to go beyond the diagonal approximation and include the leading off-diagonal terms. In that case the WDM noise correlation matrix will be banded (striped) and diagonally dominant. Importantly, the quantity that appears in the likelihood function is the inverse of the noise correlation matrix, and in general, the inverse of a banded matrix is dense, not banded. However, the inverse will be  band dominant, with exponential decay away from the diagonal~\cite{doi:10.1137/S0040585X97985224,Boito_2025}. The wavelet domain covariance matrix is real, symmetric, banded, and diagonal dominant, and a number of fast methods exist to compute approximate banded inverses for matrices of this type~\cite{10.1108/02644409910266485,doi:10.1137/0912058}. For example, if we compress the time and frequency indices $n$, $m$ into a single index $k = n + m N_t$, the WDM noise covariance matrix can be written as
\begin{equation}
C_{ij} = D_{ij} + O_{ij} = \delta_{ij} S_i + O_{ij} \, ,
\end{equation}
where $D_{ij} = \delta_{ij} S_i$ is a diagonal matrix and $O_{ij}$ is a banded matrix with $O_{ii} = 0$. A diagonal dominant matrix has $c_{ij} = D_{ik}^{-1} O_{kj} = O_{ij}/\sqrt{S_i S_j}$ with Frobenius norm $||c_{ij}|| \ll || \delta_{ij}||$. We recognize $c_{ij}$ as the off-diagonal elements of the noise {\em correlation} matrix. To leading order, the inverse noise covariance matrix is given by
\begin{equation}\label{inverse}
C^{-1}_{ij} \simeq D^{-1}_{ij} -  D^{-1}_{il} O_{lk}D^{-1}_{kj} =  \frac{\delta_{ij} S_i - O_{ij}}{S_i S_j} \, .
\end{equation}
In many applications the approximate inverse (\ref{inverse}) will be sufficient. The approximation can be improved via additional iterations of the Newton-Schulz method~\cite{https://doi.org/10.1002/zamm.19330130111,TOUTOUNIAN2013671}. Note that this approximation can be computed without the need for any expensive matrix operations. The computational cost of the likelihood calculations will be increased from the diagonal case by a factor equal to the number of stripes that are kept in $O_{ij}$. Based on our earlier calculations, of order ten stripes will be needed to account for the most significant off-diagonal terms.

We have seen that the off-diagonal elements of $C_{ij}$ can be computed from gradients of the dynamic spectral model $S(f,t)$. The expressions given in equations (\ref{freq}) and (\ref{time}) are valid when either $S(f)$ or $S(t)$. Presumably, the general expression will also include cross terms such as $\partial_t \partial_f S(f,t)$.  It may be possible to generalize the calculations presented here to cover the case of locally stationary noise using the expressions given in Ref.~\cite{Dechant_2015}.\\

 \section{Conclusions and Future Work}
 
The discrete wavelet domain provides a convenient basis to perform gravitational wave data analysis with an interpretable noise correlation matrix. The diagonal terms of the matrix encode the dynamic power spectrum $S(f,t)$, while the off-diagonal terms depend on the derivatives of $S(f,t)$. So long as the change in $S(f,t)$ across a wavelet pixel is small, it is a good approximation to treat the wavelet domain noise covariance matrix as diagonal. To ensure greater accuracy, a few off-diagonal stripes of the banded noise covariance matrix can be included at modest additional computational cost.
	 
In future work we plan to investigate the impact of non-stationary noise on long duration signals such as binary neutron star mergers in next generation ground based detectors and low mass systems detected by LISA, building upon the work in Ref.~\cite{Digman:2022igm}, and extending the likelihood to include off-diagonal terms in the noise covariance matrix.

It would also be interesting to investigate further what wavelet window function is optimal for the kinds of non-stationary noise we are likely to encounter in gravitational wave data analysis. The WDM wavelet family can be generalized to use something other than the Meyer window functions. The construction of these ``WDX'' wavelets follows that of the WDM basis, only the form of the window function ``X'' is changed. For example, the window functions can be chosen to have exponential decay in both time and frequency~\cite{Daubechies:1991wv}. The WDX transforms still result in a binary wavelet wave packet decomposition, but the balance between compactness in time and compactness in frequency can be tuned through the window choice.

 \section*{Acknowledgments}
This work was supported by the Simons Foundation award SFI-MPS-BH-00012593-04, the NASA LISA Preparatory Science Grant 80NSSC19K0320, and NSF awards PHY2513363. 

\bibliography{refs}

\end{document}